\documentstyle[epsf,onecolumn]{mn}
\title{Lensing magnification effects on the cosmic shear statistics}
\author[Takashi Hamana]{Takashi Hamana\\
Institut d'Astrophysique de Paris, CNRS, 98bis Boulevard Arago, F75014
PARIS, France\\
Max-Planck-Institut f\"ur Astrophysik,
Karl-Schwarzschild-Strasse 1, D--85748 Garching, Germany}
\date{Accepted ....; Received ....; in original form ....}
\pagerange{\pageref{firstpage}--\pageref{lastpage}}
\pubyear{....}
\begin{document}
\maketitle
\label{firstpage}

%%%%%% ABSTRACT %%%%%%%%%%%%%%%%%%%%%%%%%%%%%%%%%%%%%%%%%%%%%%%%%%%%%%
\begin{abstract}
Gravitational lensing causes a correlation between a
population of foreground large-scale structures and the observed
number density of the background distant galaxies as a consequence of
the flux magnification and the lensing area distortion.
This correlation has not been taken into account in calculations
of the theoretical predictions of the cosmic shear statistics but may
cause a systematic error in a cosmic shear measurement.
We examine its impact on the cosmic shear statistics using the
semi-analytic approach.
We find that the lensing magnification has no practical influence on 
the cosmic shear variance.
Exploring possible shapes of redshift distribution of source
galaxies, we find that the relative amplitude of the effect on the
convergence skewness is 3\% at most.
\end{abstract}

%%%%% Sec 1, INTRODUCTION %%%%%%%%%%%%%%%%%%%%%%%%%%%%%%%%%%%%%%%%%%
\section{Introduction}
The cosmic shear---coherent distortions in distant galaxy images due
to the weak lensing by large-scale structures---is now recognized as a
powerful tool to measure the mass distribution in the universe as well
as a promising way to measure the cosmological parameters (Mellier
1999; Bartelmann \& Schneider 2001 for reviews). 
Although its signal is very weak, recent reports on the detections
demonstrate that a well developed data analysis algorithm has been
established (Van Waerbeke et al.~2000; Wittman et al.~2000;
Bacon, Refregier \& Ellis 2000; Kaiser, Wilson \& Luppino 2000; Maoli
et al.~2001).

So far, the detections were obtained from relatively small fields,
which limit the statistical analyses of the surveys to second order
statistics (the variance or two-point correlation function of the 
cosmic shear).
Their amplitude reflects the amplitude of the density contrast, and
thus provides a  constraint on the combination of the values of
$\Omega_{\rm m}$ and $\sigma_8$ (Bernardeau, Van Waerbeke \& Mellier
1997, hereafter BvWM97; Jain \& Seljak 1997, Maoli et al. 2001).
The skewness of the lensing convergence is, on the other hand, known to be
sensitive to $\Omega_{\rm m}$  almost independently on $\sigma_8$
(BvWM97; Van Waerbeke et al.~2001b, hereafter vWHSCB01),
and thus it may break the degeneracy among the cosmological parameters
constrained from the second order statistics.
A precise measurement of the skewness is, therefore, one of main goals
of on-going wide field cosmic shear surveys such as the DESCART
project\footnote{For more information
about DESCART project, see http://terapix.iap.fr/Descart/.}.

It was pointed out by Hu \& Tegmark (1999) that such wide field cosmic
shear surveys have a  potential for probing the cosmological models as
accurately as the cosmic microwave background.
In order to get their ability to the fullest, there are, however, some
issues which should be developed/examined in details:
(i) making an accurate theoretical prediction of the cosmic shear
statistics at intermediate scales (0.5-10 arcmin). 
On such scales the signals will be detected easily, 
however neither perturbation theory nor the hierarchical ansatz apply 
(e.g., Jain \& Seljak 1997; vWHSCB01).
(ii) examining possible corrections which arise from higher order
correction terms in calculations of the theoretical
predictions, for example, Born approximation and lens-lens couplings 
(BvWM97; Schneider et al.~1998; vWHSCB01), and the source clustering
(Bernardeau 1998; Hamana et al.~2000).
They are especially important for the convergence skewness because it
is (in the perturbation theory sense) a quantity of 4th order of
$\delta^{(1)}$ (see \S2 for details). 
(iii) examining the impact of the intrinsic shape correlation of source
galaxies; in the theoretical calculations of the cosmic shear
statistics, it is supposed to be negligible (Croft \& Metzler 2000; 
Heavens, Refregier \& Heymans 2000; Catelan, Kamionkowski \& Blandford
2001; Crittenden et al.~2000a; 2000b).
(iv) developing a robust way to correct defects in
instruments, in particular the point spread function anisotropy (Kaiser
Squires \& Broadhurst 1995; Kuijken 1999; Erben et al.~2001; Bacon et
al.~2000).

In this paper, we focus on the lensing magnification effects on the
cosmic shear statistics, which have not been taken into consideration
in the theoretical calculation of the cosmic shear statistics but may
cause a systematic error in their measurements.
The lensing magnification has two effects on a deep galaxy survey: 
One is the flux magnification; the lensing changes the apparent galaxy
size, leaving the surface brightness invariant, therefore the flux of a
distant galaxy is magnified\footnote{We should here 
note that the lensing causes not only magnification
($\mu>1$ with $\mu$ denoting the magnification factor) but also 
demagnification ($\mu<1$).
Throughout this paper, following the usual convention,
we use the term {``magnification''} irrespective of the value of the 
magnification factor.}.
The other is the area distortion; the lensing also changes
the unit solid angle at the source plane, and thus the number
density of the distant galaxies varies with the direction in the sky
even if the intrinsic source distribution is uniform.
As a consequence, the lines of sight to distant
sheared galaxies may not be random lines of sight in presence of the
lensing magnification.
In fact, the lensing magnification effects are prominent within the
galaxy cluster region where the number density of distant
galaxies in the optical or near $IR$ bands is measured to be smaller than
the average value.
Furthermore, the variation of the galaxy number density as a function of
the distance from the cluster center (so-called, depletion curve) were
measured in some distant clusters of galaxies (e.g., Broadhurst,
Taylor \& Peacock 1995; Fort, Mellier \& Dantel-Fort 1997;
see also Mellier 1999 for a review). 
These observational facts indicate that the lensing magnification
induces an (anti-)correlation
between the distribution of the distant galaxies in the sky and the
population of the lensing structures, 
i.e., the number density of the galaxies behind a
lensing structure tends to be smaller than the average number density,  
whereas that behind a void tends to be larger than the average value.

The purpose of this paper is to quantitatively examine the lensing
magnification effects on the cosmic shear statistics, especially on
the convergence skewness.
To do this, we use the nonlinear semi-analytic approach, i.e., 
the perturbation theory approach combined with the nonlinear fitting
formula of the density power spectrum 
(Jain \& Seljak 1997; Hamana et al.~2000; vWHSCB01).
We focus on correction terms which arise from the
presence of the lensing magnification, and are not concerned with
other correction terms due to, e.g., the lens-lens coupling (vWHSCB01)
and the source clustering (Hamana et al.~2000).

The outline of this paper is as follows:
The calculations of the moments of the lensing convergence in presence
of the lensing magnification are made in \S2.
In \S3, the effect on the convergence skewness is quantitatively
examined in three cold dark matter (CDM) models with realistic models
of the redshift distributions of the source galaxy.
We conclude in \S4.

%%%%% Sec 2, Semi-analytic approach %%%%%%%%%%%%%%%%%%%%%%%%%%%%%%%%%%%%%%%%%%
\section{The semi-analytic approach}

%%%%% Sec 2-1 %%%%%%%%%%%%%%%%%%%%%%%%%%%%%%%%%%%%%%%%%%
\subsection{Fluctuation in a galaxy number count due to the lensing
magnification}

Let $n_s(>S,z)$ be the unlensed number density of galaxies with
redshift within $\Delta z$ of $z$ and with flux larger than $S$.
Then, at an angular position $\bmath{\phi}$ where the lensing
magnification is $\mu(z,\bmath{\phi})$, the number counts are
changed by the lensing
magnification effects as (e.g., Bartelmann \& Schneider 2001),
\begin{eqnarray}
\label{eq:lensed-ns}
n_s^{\rm obs}(>S,z,\bmath{\phi}) = {1\over {\mu(z,\bmath{\phi})}}
n_s\left(>{S\over{\mu(z,\bmath{\phi})}},z\right)\,.
\end{eqnarray}
Supposing that the number counts of the faint galaxies can be
approximated by a power-law over a wide range of fluxes i.e., 
$n_s(>S,z) = n(z)S^{-\alpha(z)}$, the lensed number counts are rewritten
as,
\begin{eqnarray}
\label{eq:lensed-ns2}
n_s^{\rm obs}(>S,z,\bmath{\phi})
=n_s\left(>S,z\right)\mu(z,\bmath{\phi})^{\alpha(z)-1}\,.
\end{eqnarray}
The magnification factor relates to the lensing convergence ($\kappa$)
and shear ($\gamma$) by $\mu^{-1}=|(1-\kappa)^2-\gamma^2|$.
We now rewrite the lensed source counts as
$n_s^{\rm obs}(>S,z,\bmath{\phi})=n_s(>S,z)[1+\delta
n_s(>S,z,\bmath{\phi})]$.
Taking advantage of the power-law form of the number counts and also
taking the weak lensing approximation ($\kappa\ll 1$, $\gamma\ll 1$), 
the fluctuation in the number counts due to the lensing magnification
is given by,
\begin{eqnarray}
\label{eq:delta-ns}
\delta n_s(z,\bmath{\phi})&=&\mu(z,\bmath{\phi})^{\alpha(z)-1}-1\nonumber\\
&\simeq& 2[\alpha(z)-1]\kappa_s(z,\bmath{\phi})\,,
\end{eqnarray}
where $\kappa_s(z,\bmath{\phi})$ is the lensing convergence at an
angular position $\bmath{\phi}$ for a source with redshift $z$
(Mellier 1999; Bartelmann \& Schneider 2001),
\begin{eqnarray}
\label{kappa1}
\kappa_s(z,\bmath{\phi})={{3 \Omega_{\rm m}} \over 2} {{H_0}\over c}
\int_0^{\chi_s(z)}d \chi\, g(\chi,\chi_s)
\delta(\chi,\bmath{\phi})\,,
\end{eqnarray}
where $g$ is the so-called lensing efficiency function defined by,
\begin{eqnarray}
\label{g}
g(\chi,\chi_s)={{H_0}\over c}{{f(\chi)f(\chi_s-\chi)} \over
{f(\chi_s)a(\chi)}}\,.
\end{eqnarray}
Here $\chi$ denotes the radial comoving distance, $a$ is the scale
factor normalized by its present value, and $f(\chi)$ denotes
the comoving angular diameter distance,
defined as $f(\chi)=K^{-1/2} \sin K^{1/2} \chi$, $\chi$, $(-K)^{-1/2}
\sinh (-K)^{1/2} \chi$ for $K>0$, $K=0$, $K<0$, respectively, where
$K$ is the curvature which can be expressed as
$K=(H_0/c)^2(\Omega_{\rm m}+\Omega_\lambda-1)$.
The lensing convergence is therefore the projected density contrast
weighted by the distance combination and the scale factor along the line of
sight to a source.
Note that, in the expression (\ref{eq:delta-ns}), the fluctuation does
not depend on the flux because of the power-law form of the number
counts.
In the following sections, we will therefore not explicitly denote
the flux dependence of the source counts.

%%%%% Sec 2-2 %%%%%%%%%%%%%%%%%%%%%%%%%%%%%%%%%%%%%%%%%%
\subsection{Cosmic shear statistics in the presence of the lensing
magnification}

Let us consider the measured convergence that results from averages
made over many distant galaxies located at different distances.
Denoting the smoothing scale by $\theta$, such an average can formally
be written as,
\begin{eqnarray}
\label{average-kappa}
\kappa_\theta={
{\sum_{i=1}^{N_s} W_\theta(\bmath{\phi}_i)
\kappa_s(z_i,\bmath{\phi}_i)}
\over
{\sum_{i=1}^{N_s} W_\theta(\bmath{\phi}_i)}
}\,,
\end{eqnarray}
where $W_\theta(x)$ denotes the weight function of the average,
$N_s$ is the number of source galaxies, and $\bmath{\phi}_i$ and $z_i$ 
are the direction and redshift of $i$-th source, respectively.
For the weight function, the angular top-hat filter (BvWM97) and/or
compensated filter (Schneider et al.~1998) are commonly adopted (e.g.,
Van Waerbeke et al.~2001b).
In what follows, we consider the top-hat filter for the weight
function, and in this case equation (\ref{average-kappa}) is reduced to
$\kappa_\theta=\sum_{i=1}^{N_s^j} \kappa_s(z_i,\bmath{\phi}_i)/N_s^j$,
where $N_s^j$ is the number of source galaxies within an aperture $\theta$
centered on a direction $\bmath{\phi}_j$.
Taking the continuous limit for the source distribution\footnote{See
Bernardeau (1998) for a discussion on this point.}, equation
(\ref{average-kappa}) can be rewritten by,
\begin{eqnarray}
\label{average-kappa2}
\kappa_\theta={
{\int d^2 \phi\, W_\theta(\bmath{\phi}) \int_0^{\chi_H} d \chi\, 
\kappa_s(\chi,\bmath{\phi}) n_s^{\rm obs}(\chi,\bmath{\phi})}
\over
{\int d^2 \phi\, W_\theta(\bmath{\phi}) \int_0^{\chi_H} d \chi\,
n_s^{\rm obs}(\chi,\bmath{\phi})}
}\,,
\end{eqnarray}
where $\chi_H$ is the distance to the horizon and $n_s^{\rm
obs}(\chi,\bmath{\phi})$ is the source number count defined by
equation (\ref{eq:delta-ns}) which effectively describes the redshift
distribution of the source. 

In what follows, the distance distribution of the unlensed number
counts is supposed to be
normalized to unity, $\int_0^{\chi_H}d\chi\, n_s(\chi)=1$.
Let us now expand equation (\ref{average-kappa2}) in terms of
$\delta$ using the perturbation theory approach (BvWM97).
The presence of the lensing magnification does not change the
expression of the first order term,
\begin{eqnarray}
\label{k1}
\kappa_\theta^{(1)} &=& {{3 \Omega_{\rm m} H_0} \over {2c}} 
\int d^2 \phi\, W_\theta(\bmath{\phi})
\int_0^{\chi_H} d \chi\, n_s(\chi)
\int_0^{\chi} d \chi'\,
g(\chi',\chi) \delta^{(1)}(\chi',\bmath{\phi})\,.
\end{eqnarray}
The second order convergence consists of two terms:
One comes from the second order density perturbation, which is
formally written by replacing the subscript $^{(1)}$ in the first order
expression (\ref{k1}) with $^{(2)}$ (BvWM97).
The other is due to the lensing magnification, 
\begin{eqnarray}
\label{kmag}
\kappa_\theta^{{\rm mag.}(2)}
&=& 2 \left( {{3 \Omega_{\rm m} H_0} \over {2c}} \right)^2
\int d^2 \phi\, W_\theta(\bmath{\phi})
\int_0^{\chi_H} d \chi\, n_s(\chi) [\alpha(\chi)-1]
\int_0^{\chi} d \chi'\,
g(\chi',\chi) \delta^{(1)}(\chi',\bmath{\phi})
\int_0^{\chi} d \chi''\,
g(\chi'',\chi) \delta^{(1)}(\chi'',\bmath{\phi})\nonumber\\
&&-\kappa_\theta^{(1)} 
{{3 \Omega_{\rm m} H_0} \over {c}} 
\int d^2 \phi\, W_\theta(\bmath{\phi})
\int_0^{\chi_H} d \chi\, n_s(\chi) [\alpha(\chi)-1]
\int_0^{\chi} d \chi'\,
g(\chi',\chi) \delta^{(1)}(\chi',\bmath{\phi})\,.
\end{eqnarray}
Using the small angle approximation (Kaiser 1992),
equation (\ref{k1}) is rewritten in terms of the Fourier transform of
the density contrast, $\delta(k)$, by,
\begin{eqnarray}
\label{k1f}
\kappa_\theta^{(1)} &=& {{3 \Omega_{\rm m} H_0} \over {2c}} 
\int_0^{\chi_H} d \chi\, n_s(\chi)
\int_0^{\chi} d \chi'\,g(\chi',\chi) 
\int {{d^3 k} \over {(2\pi)^3}} W[f (\chi') k_\perp
\theta] \delta^{(1)}[\bmath{k};\chi'] \exp[i k_{\chi} f(\chi')]\,,
\end{eqnarray}
where the wave vector $\bmath{k}$ is decomposed into the line-of-sight
component $k_\chi$ and perpendicular to it, $\bmath{k_\perp}$, and
$W(x)$ is the Fourier transform of the weight function.
In the case of the top-hat filter, $W(x)=2J_1(x)/x$ where $J_1$ is
the spherical Bessel function. 
In the same manner, equation (\ref{kmag}) is rewritten by,
\begin{eqnarray}
\label{kmagf}
\kappa_\theta^{{\rm mag.}(2)}
&=& 2 \left( {{3 \Omega_{\rm m} H_0} \over {2c}} \right)^2
\int_0^{\chi_H} d \chi\, n_s(\chi) [\alpha(\chi)-1]
\int_0^{\chi} d \chi'\,g(\chi',\chi)
\int_0^{\chi} d \chi''\,g(\chi'',\chi) \nonumber\\
&&\times 
\int {{d^3 k'} \over {(2\pi)^3}} \delta^{(1)}[\bmath{k'};\chi'] 
\exp[i {k'}_{\chi} f(\chi')]
\int {{d^3 k''} \over {(2\pi)^3}} \delta^{(1)}[\bmath{k''};\chi''] 
\exp[i {k''}_{\chi}f(\chi'')]
W[|f (\chi') \bmath{{k'}_\perp} + f (\chi'') \bmath{{k''}_\perp}|\theta]
\nonumber\\ 
&&-\kappa_\theta^{(1)} 
{{3 \Omega_{\rm m} H_0} \over {c}} 
\int_0^{\chi_H} d \chi\, n_s(\chi) [\alpha(\chi)-1]
\int_0^{\chi} d \chi'\, g(\chi',\chi) 
\int {{d^3 k'} \over {(2\pi)^3}} W[f (\chi') {k'}_\perp
\theta] \delta^{(1)}[\bmath{k'};\chi'] \exp[i {k'}_{\chi} f(\chi')]\,.
\end{eqnarray}

The lensing magnification effect makes the average of the convergence
non-zero,
\begin{eqnarray}
\label{ave-kap}
\langle \kappa_\theta \rangle 
&=& \langle \kappa_\theta^{{\rm mag.}(2)} \rangle\nonumber\\
&=& 2 \left( {{3 \Omega_{\rm m} H_0} \over {2c}} \right)^2
\int_0^{\chi_H} d \chi\, n_s(\chi) [\alpha(\chi)-1]
\int_0^{\chi} d \chi'\,g^2(\chi',\chi)
I_0(\chi',0)\nonumber\\
&&-2 \left( {{3 \Omega_{\rm m} H_0} \over {2c}} \right)^2
\int_0^{\chi_H} d \chi\, n_s(\chi)
\int_0^{\chi_H} d \chi'\, n_s(\chi') [\alpha(\chi')-1]
\int_0^{\chi} d \chi''\, g(\chi'',\chi) g(\chi'',\chi') 
I_0(\chi'',\theta)\,,
\end{eqnarray}
where
\begin{eqnarray}
\label{I0}
I_0(\chi,\theta)={1\over {2\pi}} \int dk\, k
W^2[f(\chi'')k\theta]P_{\rm lin} (\chi,k)\,,
\end{eqnarray}
with $P_{\rm lin} (\chi,k)$ being the linear matter power spectrum.
The variance is not affected by the lensing magnification and is given by,
\begin{eqnarray}
\label{variance}
V_\kappa (\theta)
= \langle (\kappa_\theta- \langle \kappa_\theta \rangle)^2 \rangle
=\langle {\kappa_\theta^{(1)}}^2 \rangle
= \left( {{3 \Omega_{\rm m} H_0} \over {2c}} \right)^2
\int_0^{\chi_H} d \chi\, n_s(\chi)
\int_0^{\chi} d \chi'\,g^2(\chi',\chi)
I_0(\chi',\theta)\,.
\end{eqnarray}
In presence of the lensing magnification, the skewness parameter,
defined by $S_3(\theta)=\langle (\kappa_\theta- \langle
\kappa_\theta \rangle)^3 \rangle/V_\kappa^2(\theta)$, consists of two terms:
One comes from the second order perturbation, $\langle
\kappa_\theta^3\rangle^{\rm 2PT}=3 \langle
{\kappa_\theta^{(1)}}^2 \kappa_\theta^{(2)}\rangle$ (see BvWM97 and
Hamana et al.~2000 for the explicit expression).
The other arises from the lensing magnification, 
\begin{eqnarray}
\label{s3mag}
\langle \kappa_\theta^3\rangle^{\rm mag}&=&\langle
(\kappa_\theta^{(1)}+\kappa_\theta^{{\rm mag.}(2)}-
\langle \kappa_\theta \rangle)^3 \rangle
=3 \langle {\kappa_\theta^{(1)}}^2 \kappa_\theta^{{\rm mag.}(2)}
\rangle 
-3 \langle {\kappa_\theta^{(1)}}^2 \rangle 
\langle \kappa_\theta^{{\rm mag.}(2)} \rangle  \nonumber\\
&=& 12 \left( {{3 \Omega_{\rm m} H_0} \over {2c}} \right)^4
\int_0^{\chi_H} d \chi\, n_s(\chi) [\alpha(\chi)-1]
\left[ 
\int_0^{\chi_H} d \chi'\, n_s(\chi')
\int_0^{\chi'} d \chi''\,g(\chi'',\chi)g(\chi'',\chi')
I_0(\chi'',\theta)\right]^2\nonumber\\
&&-V_\kappa(\theta) \times  12 \left( {{3 \Omega_{\rm m} H_0} 
\over {2c}} \right)^2
\int_0^{\chi_H} d \chi\, n_s(\chi)
\int_0^{\chi_H} d \chi'\, n_s(\chi') [\alpha(\chi')-1]
\int_0^{\chi} d \chi''\, g(\chi'',\chi) g(\chi'',\chi') 
I_0(\chi'',\theta)\,.
\end{eqnarray}
To derive the last expression, we used an approximation,
$\int_0^{2\pi}d\vartheta \sin\vartheta
W(|\bmath{k}+\bmath{k'}|)\simeq 2\pi W(k)W(k')$ with $\vartheta$ being
the angle between the wave vectors $\bmath{k}$ and $\bmath{k'}$.
For the top-hat window function, the error this approximation induces
is extremely weak, for instance it is less than 1\% for $n\sim-1.5$
with $n$ being the effective power-law index of the matter power spectrum
(Bernardeau 1998). 
Notice that, in the case of $\alpha(\chi)=0$, which corresponds to the
case that sources are selected on a surface brightness criterion 
(see \S 4 for discussion on this point), 
the second term reduces to $12V_\kappa^2 (\theta)$.
As this immediately suggests, the cosmology dependence in the
skewness correction term is weak, actually $\Omega_{\rm m}$ in the
coefficients of equation (\ref{s3mag}) is canceled out by that in 
$V_\kappa^2(\theta)$ (this point is demonstrated in Figure \ref{fig:s3nl}).

The above calculations are based on the perturbation theory approach.
It is well known that on sub-degree scales the nonlinearity in the
evolution of the density field is very important for the cosmic shear
statistics (Jain \& Seljak 1997; wVHSCB).
We take the nonlinearity into account in our computations in the
following way:
For the variance, the effect of nonlinear evolution of the density power
spectrum can be included by replacing the linear power spectrum with
the nonlinear  power spectrum, i.e.,
$P_{\rm lin}(a,k) \rightarrow P_{\rm NL}(a,k)$ (Jain \& Seljak 1997).
We use the fitting formula for the nonlinear power spectrum  given by
Peacock and Dodds  (1996).
For the skewness correction term, all density contrast terms needed
for its calculation, equation (\ref{kmagf}), correspond to the linear
order.  
This is the same situation as for the variance.  
Following the procedure used for this latter case, we simply replace
the linear power spectrum with the nonlinear one to include nonlinear 
effects.
We adopt the semi-analytic calculation of the skewness in the
nonlinear regime developed by vWHSCB01, which is based on the fitting
formula of the density bispectrum by 
Scoccimarro \& Couchman (2000).

It should be here noted that,  
comparing equation (\ref{ave-kap}) with (\ref{variance}), 
the amplitude of the shift in average of the convergence from
zero is found to be the same order of the variance, and thus is of
order of $O(10^{-4})$.
This shift has no practical effect on cosmic shear statistics because
the constant shift has no effect on the second and higher order
statistics by definition.

%%%%% Sec 3, Numerical results %%%%%%%%%%%%%%%%%%%%%%%%%%%%%%%%%%%%%%%%%%
\section{Numerical results}

%%%%% TABLE 1 %%%%%%%%%%%%%%%%%%%%%%%%%%%%%%%%%%%%%%%%%%%%%%%%%%%%%%%%%%%
%
\begin{table}
\caption{Cosmological parameters.}
\label{table:cosmo}
\begin{tabular}{ccccc}
\hline
Model & $\Omega_{\rm m}$ & $\Omega_\lambda$ & $h$ & $\sigma_8$ \\
\hline
SCDM & 1.0 & 0.0 & 0.5 & 0.6 \\
OCDM & 0.3 & 0.0 & 0.7 & 0.85 \\
$\Lambda$CDM & 0.3 & 0.7 & 0.7 & 0.9 \\
\hline
\end{tabular}
\end{table}

In this section, we numerically examine the lensing magnification
effect on the convergence skewness.
We take three CDM models, a flat model with ($\Lambda$CDM) and without
cosmological constant (SCDM) and an open model (OCDM).
We use the fitting formula of the CDM power spectrum given by Bond \&
Efstathiou (1984) normalized by the local galaxy cluster abundance (Eke,
Cole \& Frenk 1996; Kitayama \& Suto 1997).
The parameters in the models are listed in Table \ref{table:cosmo}.

%%%%% Table 2 %%%%%%%%%%%%%%%%%%%%%%%%%%%%%%%%%%%%%%%%%%%%%%%%%%%%%%
\begin{table}
\caption{Parameters in models of the redshift distribution of source
galaxies ($a$, $b$, $z_\ast$) and their characteristics 
(the mean redshift $\langle z \rangle$ and the root-mean-square of the
distribution $\Delta z$)}
\label{table:ns}
\begin{tabular}{cccccc}
\hline Model & $a$ & $b$ & $z_\ast$  & $\langle z \rangle$ & $\Delta z$\\ 
\hline 
A & 2 & 1.5 & 0.798 & 1.2 & 0.572\\ 
B & 3 & 1.8 & 0.813 & 1.2 & 0.456\\ 
C & 5 & 2.5 & 1.11  & 1.5 & 0.400\\ 
D & 2 & 1.5 & 0.598 & 0.9 & 0.429\\
\hline
\end{tabular}
\end{table}
%%%%%%%%%%%%%%%%%%%%%%%%%%%%%%%%%%%%%%%%%%%%%%%%%%%%%%%%%%%%%%%%%%%

We assume that $n_s(z)$ takes the form,
\begin{eqnarray}
\label{ns}
n_s(z) =  {b \over {z_\ast
\Gamma[(1+a)/b]}}  \left( {z\over {z_\ast}} \right)^a
\exp \left[-\left({z\over{z_\ast}} \right)^b \right].
\end{eqnarray}
where $\Gamma(x)$ is the Gamma function.
We explore four models for the shape of the distribution.
The parameters as well as characteristics (the mean redshift $\langle
z \rangle$ and the root-mean-square of the distribution $\Delta z$) in
each model are listed in Table \ref{table:ns}.
Note that only model A matches roughly the observed redshift
distribution of galaxies in current cosmic shear detections (Van
Waerbeke et al.~2000).  
However, we test the other models to see a variation in the
lensing magnification effect due to possible changes in the shape
of the redshift distribution.

%%%%%%%%% Figure 1 %%%%%%%%%%%%%%%%%%%%%%%%%%%%%%%%
\begin{figure}
\begin{center}
\begin{minipage}{11cm}
\begin{center} 
\epsfxsize=11cm \epsffile{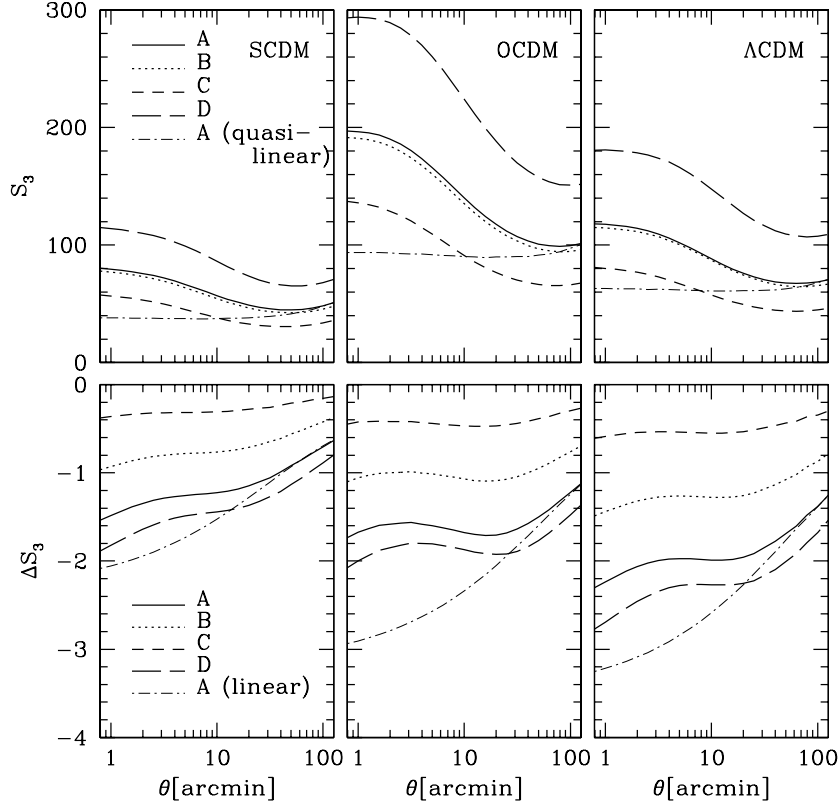}
\end{center}
\end{minipage}
\end{center}
\caption{{\it Upper panels}: 
Skewness parameter $S_3$ of the lensing convergence evaluated
without taking the lensing magnification effect into account.
The dot-dashed line is the quasi-linear perturbation theory
computation for A model. 
The other curves represent the semi-analytic computations with the
nonlinear matter bispectrum fitting formula (vWHSCB01).
{\it Lower panels}:
Skewness correction term due to the lensing magnification
effects (for a case of $\alpha(z)=0$).
The dot-dashed line is the linear theory computation for A model. 
The other lines represent the semi-analytic nonlinear computations.}
\label{fig:s3nl}
\end{figure}

The upper panels of Figure \ref{fig:s3nl} show the skewness parameter
$S_3(\theta)$ evaluated
without taking the magnification effects into account.
As the figure clearly shows, the skewness is very sensitive
to both $\Omega_{\rm m}$ and the mean redshift of sources, but insensitive
to the shape of the redshift distribution (comparing model A with B).

The lower panels of Figure \ref{fig:s3nl} show the skewness correction
due to the lensing magnification effects.
We took $\alpha(z)=0$ which gives the strongest estimate of
the lensing magnification effect as will be discussed in \S 4.
The nonlinear result deviates from the linear one on a sub-degree
scale and the nonlinearity reduces the amplitude of the magnification
effect as it has stronger
influence on $V_\kappa(\theta)$ than on $\langle \kappa_\theta^3
\rangle^{\rm mag}$.
Between the scales displayed, the skewness correction is almost constant,
the variation is less than 1.4.
The lensing magnification effect becomes stronger as the mean
redshift becomes lower and as the distribution becomes broader.
This redshift distribution dependence is, at lest qualitatively,
similar to that on the {\it source clustering effect}, which may be
due to the similarity in their phenomena (Hamana et al.~2000).
The most important point found in Figure \ref{fig:s3nl} is, however, 
that the correction term is small, 3\% correction at largest.

%%%%% Section 4 Discussion %%%%%%%%%%%%%%%%%%%%%%%%%%%%%%%%%%%%%%%%%%
\section{Discussion and conclusion}
We have examined the lensing magnification effects on the convergence
skewness using the nonlinear semi-analytic approach.
Numerical computations were done only for the case of $\alpha(z)=0$.
Does this choice have a special meaning ?
So far, little is known about the number counts of distant galaxies
with measured redshifts, in particular at $z>0.2$ where most of
the source galaxies are located.
We can, however, put a constraint on the range of the ``effective
value'' of $\alpha$ (under the assumption of no strong redshift
evolution in $\alpha$) as follows:
(i) $\alpha>0$ by definition.
(ii) The observational facts that the number density of the distant
galaxies behind the lensing galaxy clusters in optical or near $IR$
bands is smaller than that measured in the field region (e.g., Fort et
al.~1997).
This indicates that $\alpha<1$ (comes from eq.~(\ref{eq:lensed-ns2})
with $\mu>1$ within cluster region).
Therefore, it may be said that the possible range of the ``effective value''
of $\alpha$ is $0<\alpha<1$.
Within this range, $\alpha=0$ gives the upper limit of the lensing
magnification effect on the convergence skewness.
We may, therefore, conclude that the lensing magnification has no
significant effect on the convergence skewness, its amplitude could be
3\% at most.

In the above discussion, we implicitly assumed that the source
galaxies are selected by a flux threshold.
However, one can adopt the selection with a surface brightness
criterion. 
In this case, the flux magnification has no influence on the cosmic
shear measurements, but the area magnification still exists.
Therefore, the lensing magnification effect in this case can be
estimated by setting $\alpha=0$.
Thus, the effect is stronger for the surface brightness selection than
for the flux selection.

Finally, we notice a limitation of our calculation.
We used the weak lensing approximation to derive equation (\ref{eq:delta-ns}).
This approximation works well except for very rare directions such like
the core of clusters of galaxies where the lensing surface mass
density is very high.
Such regions where a strong lensing event can take place are very small,
$\theta\la 0.5$ arcmin, at largest.
Therefore, it is expected that the strong lensing may change the
above results on scales below 1 arcmin.

%%%%% Acknowledgments %%%%%%%%%%%%%%%%%%%%%%%%%%%%%%%%%%%%%%%%%%
\section*{Acknowledgments}
The author would like to thank F.~Bernardeau for discussions,
M.~Bartelmann for a careful reading of the manuscript, and
L.~Van Waerbeke for providing the FORTRAN code
to compute the nonlinear skewness of convergence. 
He would also like to thank an anonymous referee for constructive
comments which improved the presentation of this paper.
This research was supported in part by the Direction de la Recherche
du Minist{\`e}re Fran{\c c}ais de la Recherche.   
Numerical computations in this work was partly carried out at
the TERAPIX data center at IAP.

%%%%%% REFERENCE %%%%%%%%%%%%%%%%%%%%%%%%%%%%%%%%%%%%%%%%%%%%%%%%%%%%

\bsp % ``This paper has been produced using the ...''
\label{lastpage}

\end{document}